\documentclass[useAMS,usenatbib]{mn2e}
\usepackage{amsmath}
\usepackage{times}
\usepackage{psfig}
\usepackage{graphics,graphics}
\usepackage{ls cape}
\usepackage{epsf}
\usepackage{epsfig}
\usepackage{color}
\usepackage{dcolumn}
\usepackage{natbib}
\bibpunct{(}{)}{;}{a}{}{,}

\newcommand{\mdot}{\mbox{$\dot{M}$}}
\newcommand{\Rstar}{\mbox{$R_\ast$}}

\newcommand{\myr}{\mbox{$M_\odot\,{\rm yr}^{-1}$}}
\newcommand{\lsim}{\raisebox{-.4ex}{$\stackrel{<}{\scriptstyle \sim}$}}
\newcommand{\msim}{\raisebox{-.4ex}{$\stackrel{>}{\scriptstyle \sim}$}}

\def \etal   {\hbox{et~al.\/}}
\def\changed{}
\def\nchanged{}

\title[Clumped stellar winds in HMXBs]{Clumped stellar winds in 
supergiant high-mass X-ray binaries:  X-ray variability and photoionization}
\author[Oskinova,  Feldmeier \& Kretschmar]
{L. M. Oskinova$^{1}$,\thanks{E-mail: lida@astro.physik.uni-potsdam.de},  
A. Feldmeier$^{1}$, P. Kretschmar$^2$ \\
$^{1}$ Institute for Physics and Astronomy, University of Potsdam,
14476 Potsdam, Germany\\
$^2$  European Space Astronomy Centre (ESA/ESAC),
  Science Operations Department,
  Villanueva de la Cañada (Madrid), Spain}

\begin{document}

\date{Accepted . Received ; in original form 21.02.2011 22:32}

\pagerange{\pageref{firstpage}--\pageref{lastpage}} \pubyear{2005}

\maketitle

\label{firstpage}

\begin{abstract}
The clumping of massive star winds is an established paradigm
confirmed by multiple lines of evidence and supported by stellar
wind theory.  The purpose of this paper is to bridge the gap between
detailed models of inhomogeneous stellar winds in single stars and
the phenomenological description of donor winds in supergiant
high-mass X-ray binaries (HMXBs). We use results from time-dependent
hydrodynamical models of the instability in the line-driven wind of a
massive supergiant star to derive the time-dependent accretion rate
onto a compact object in the Bondi-Hoyle-Lyttleton approximation.  The
strong density and velocity fluctuations in the wind result in strong
variability of the synthetic X-ray light curves. The model predicts a
large scale X-ray variability, up to eight orders of magnitude, on
relatively short timescales. The apparent lack of evidence for such
strong variability in the observed HMXBs indicates that the details of
accretion process act to reduce the variability due to the stellar
wind velocity and density jumps.

We study the absorption of X-rays in the clumped stellar wind by means
of a 2-D stochastic wind model. The monochromatic absorption in cool
stellar wind in dependence on orbital phase is computed for realistic
stellar wind opacity.  We find that absorption of X-rays changes
strongly at different orbital phases. The degree of the variability due 
to the absorption in the wind depends on the shape of the wind clumps and 
is stronger in case of oblate clumps.

We address the photoionization in the clumped wind, and show that the
degree of ionization is affected by the wind clumping. A correction
factor for the photoionization parameter is derived. It is shown that
the photoionization parameter is reduced by a factor ${\cal X}$ 
compared to the smooth wind models with the same mass-loss rate, where 
${\cal X}$ is the wind inhomogeneity parameter. We conclude that wind 
clumping must also be taken into account when comparing the observed and 
model spectra of the photoionized stellar wind. \end{abstract}


\section{Introduction}
Massive luminous OB-type stars possess strong stellar winds. The winds
are fast, with typical velocities up to 2500\,km\,s$^{-1}$, and dense,
with mass-loss rates $\dot{M}\msim 10^{-7}$\,\myr. The driving
mechanism for the mass-loss from OB stars has been identified with
radiation pressure on spectral lines \citep[][CAK]{cak1975}. Early on
\citet{ls1970} suggested that the stationary solution for a
line-driven wind is unstable leading to shocks in stellar winds.
\citet{lw1980} proposed that the winds break up into a population of
dense blobs. \citet{ow1988} performed numerical simulations of the
nonlinear evolution of wind instabilities and showed that even
small-amplitude perturbations in the radial velocity near the wind
base result in a highly structured wind, with relatively slow, dense
shells separated from regions of high-speed, rarefied flow by strong
shocks. \citet{feld1997a, feld1997b} used hydrodynamic simulations to
model the evolution of wind instabilities and predicted the X-ray
emission from single stars matching the observed X-ray fluxes. These
non-stationary hydrodynamic simulations are used in this
paper. \citet{ro2002} studied the evolution of wind structures and
demonstrated that the winds can remain inhomogeneous out to distances of
1000\,R*, while \citet{do2005} presented first attempts of 2-D radiative
hydrodynamic models of stellar winds. 

These theoretical predictions of stellar wind clumping are confirmed
by observations. From analysis of wind lines in the UV spectra
\citet{lu1982} postulated the existence of non-monotonic velocity
fields in the winds.  \citet{hil1991} induced the existence of wind
clumping from the shape of electron-scattering wings in emission lines
of Wolf-Rayet (WR) stars. Stochastic variable structures in the
He\,{\sc ii} $\lambda 4686$\,\AA\ emission line in an O-supergiant
were found by \citet{ev1998}, and explained as excess emission from
the wind clumps. \citet{mar2005} investigated the line-profile
variability of H$\alpha$ for a large sample of O-type
supergiants. They concluded that the observed variability can be
explained by a wind model consisting of coherent or broken
shells. \citet{lep1999, lep2008} presented direct spectroscopic
evidence of clumping in O and WR star winds. \citet{pm2010}
established spectroscopic signatures of the wide-spread existence of
wind clumping in B supergiants.

The inhomogeneity affects stellar wind diagnostics. \citet{sf1981}
used \emph{Einstein} spectra of O-stars to study the transfer of
X-rays through a uniform stellar wind. By matching the model and the
observed X-ray spectrum of the O-type supergiant $\zeta$ Pup they
found that the mass-loss rate derived from X-ray spectra is lower by
factor of a few than the mass-loss rates obtained from fitting the
H$\alpha$ line as well as from observations at radio and IR
wavelengths.  As most plausible explanation for this discrepancy they
suggested the neglect of the clumping in O-star winds. \citet{wc2001}
proposed that a clumpy wind structure may help to explain the observed
X-ray emission line profiles in the spectra of O-supergiants. A theory 
of X-ray transfer in clumped wins that accounts for clumps of any 
shapes and optical depths was first developed by \citet{feld2003,osk2004}.
Indeed, accounting for wind clumping allows to consistently model
the UV/optical and the X-ray spectra of O-stars \citep{osk2006}. 

Presently, clumping in first approximation (``microclumping'') is
taken into account in all up-to-date stellar wind codes.  This
approximation assumes that clumps are optically thin and that the
interclump medium is void.  Microclumping leads to smaller empirical
mass-loss rates when using diagnostics that depend on processes
scaling with density squared ($\rho^2$), such as H$\alpha$ and other
emission  lines, or radio free-free emission
\citep{hil1991, ham1998}. \citet{bour2005} and \citet{ful2006}
demonstrated that mass-loss rates must be reduced by orders of
magnitude in order to reproduce the UV-resonant lines (which depend on
density linearly) simultaneously with the $\rho^2$ based
diagnostics \footnote{\citet{wc2010} pointed out that the extreme UV
  radiation may be important for the mass-loss diagnostics based on
  resonant lines.}.  These drastic reductions of empirically deduced
mass-loss rates would have strong consequences for the stellar
evolution and stellar feedback models.  

A way to solve this problem was suggested by \citet{osk2007}. It
was found that waiving the microclumping approximation i.e.\ allowing 
for clumps of arbitrary optical depth (``macroclumping'') in the 
spectral models eliminates the need to strongly reduce stellar mass-loss 
rates.  It was shown that the strength of optically thick diagnostic lines, 
such as the P\,{\sc  v} resonance line in OB star spectra, is reduced by 
macroclumping effects, while the strength of optically thin lines, such 
as H$\alpha$, is not affected. Therefore, accounting for macroclumping in 
analyzing stellar spectra yields empirically derived mass-loss rates that 
are only a factor 2\,--\,3 lower than the ``classical'' CAK mass-loss
rates. These conclusions are confirmed by 2D model studies 
\citep{sun2010} and full 3D radiative-transfer models 
(\v{S}urlan \etal, submitted).

The CAK mass-loss rates of massive stars are in good agreement with
observed X-ray fluxes from high-mass X-ray binaries (HMXBs). HMXBs are
the products of binary star evolution \citep[e.g.][]{shklov1967,iben1995}. 
These systems consist  of an early-type massive star and a compact companion, 
neutron star (NS) or black hole (BH), on a close orbit. Accretion of 
stellar  wind onto the compact companion powers the high X-ray luminosity of
$\sim 10^{33}$\,--\,$10^{39}$\,erg\,s$^{-1}$. HMXBs can be divided
in subclasses depending on the spectral type of the donor star --
supergiant (SG) or Be-star, and the type of variability -- persistent or
transient.

In this paper we consider SG HMXBs. The basic physics of neutron-star
accretion in stellar winds was described in the seminal paper by
\citet{do1973}, who suggest that the Bondi-Hoyle-Lyttleton accretion
powers the X-ray luminosity in HMXBs. \citet{kal1982} presented
theoretical models for the temperature and ionization structure of
spherically symmetric, constant density, gaseous nebulae surrounding
compact X-ray sources.  Two-dimensional hydrodynamic simulations of
the gas flow in the orbital plane of a massive X-ray binary system
were presented by \citet{blond1990}. The simulations revealed the
presence of dense filaments of compressed gas formed in the non-steady
accretion bow shock and wake of the compact object. The underlying
assumption in all these models is an initially smooth donor
stellar wind with the CAK density distribution. In a recent review,
\citet {edgar2004} demonstrated that despite the simplifications made,
Bondi-Hoyle-Lyttleton predictions for the accretion rate are quite
accurate. \citet{neg2010} reviewed recent works on stellar wind
accretion, and suggested that the real sources differ from the
idealized situation of Bondi-Hoyle-Lyttleton accretion in three main
ways: the winds of massive stars are highly structured; accretion to a
compact object is an unstable process; and the magnetic field of the
neutron star may affect the flow of material. In this paper we
investigate the immediate consequences of stellar wind clumping for
the accretion rate, wind ionization, and the absorption of X-rays in
the unperturbed portion of the wind.

The observational evidence of clumped accretion flow in SG HMXBs is
growing. \citet{sako2003} reviewed spectroscopic results obtained by
X-ray observatories for several wind-fed HMXBs. They concluded that
the observed spectra can be explained only as originating in a clumped
stellar wind where cool dense clumps are embedded in rarefied
photoionized gas. \citet{vdm2005} studied the X-ray light curve and
spectra of the HMXB 4U\,1700-37. They showed that the feeding of the NS by
a strongly clumped stellar wind is consistent with the observed
stochastic variability.


\begin{table}
\begin{center}
\caption{Stellar model parameters (see the full set of parameters in 
\citet{feld1997a})}
\vspace{1em}
\renewcommand{\arraystretch}{1.2}
\begin{tabular}[h]{lcc}  \hline
\hline
Spectral type & & O9.7Ib \\
Mass & $M_\ast$ & 34\,$M_\odot$ \\
Radius & $R_\ast$ & 24\,$R_\odot$ \\
Terminal speed & $v_\infty$ & 1850\,km/s \\
Mass loss rate & $\dot{M}$ & $3\times 10^{-6}$\,\myr \\
\hline \hline
\end{tabular}
\label{tab:starmod}
\end{center}
\end{table}

\citet{int2005} proposed that the flaring behavior of supergiant fast
X-ray transients (SFXT) can be explained by the accretion of wind
blobs. This suggestion was corroborated by the studies of hard X-ray
flares and quiescent emission of SFXT systems by \citet{wal2007}, who
estimated wind clump parameters and found them to be in agreement with
the macroclumping paradigm of stellar winds. \citet{neg2008} presented
a common framework for wind accreting sources, in the context of
clumpy wind models. They postulated that the different flaring behavior
can be an immediate consequence of diverse orbital geometries.

\begin{figure*}
\centering
\includegraphics[width=15cm]{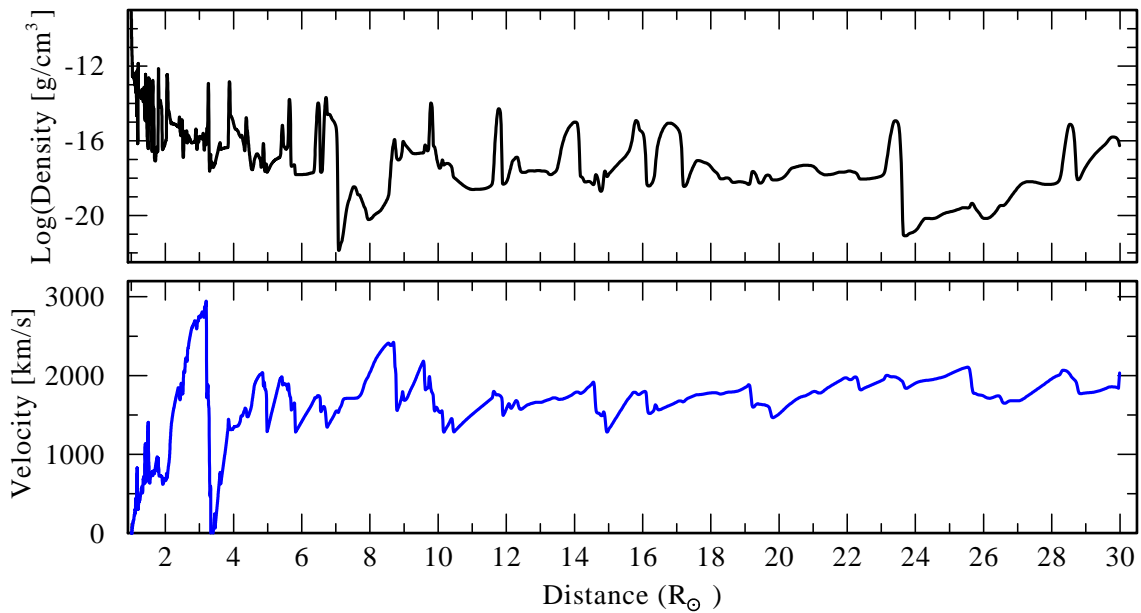}
\caption{ A snapshot of the density (upper panel) and the
velocity (lower panel) structure of the stellar wind in a
O9.5 supergiant star as predicted by 
time-dependent hydrodynamic simulations \citep{feld1997a,feld1997b}.
The upper panel shows that the density fluctuations in the wind  
have many orders of magnitude, while the lower panel shows the 
strong wind velocity jumps.}
\label{fig:rhov}
\end{figure*}
%

\citet{krey2008} studied the temporal variability of Vela\,X-1.  They
interpreted flares and off states as being due to a strongly
structured wind of the optical companion. They suggested that when the
NS in Vela X-1 encounters a cavity with strongly reduced density, the
X-ray flux drops, thus triggering the onset of the propellbibdefinitions.texer effect, which
inhibits further accretion (off states). From the
analysis of the flaring behavior of Vela\,X-1  \citet{furst2010} concluded
that a mixture of a clumpy wind, shocks, and turbulence can explain
the flare brightness distribution. {\changed \citet{sid2010} suggested that 
the X-ray variability of the HMXB IGR\,J08408--4503 may be explained by 
 stellar wind clumping.} \citet{boz2011} observed a bright
X-ray flare in the HMXB IGR\,J18410-0535.  They concluded that the flare
was produced by the accretion of massive clump onto the
compact object hosted in this system.

\citet{duc2009} developed a phenomenological stellar wind model for OB
supergiants to investigate the effects of accretion from a clumpy wind
on the luminosity and variability properties of high-mass X-ray
binaries. They concluded that the model can reproduce the observed
light curves well. They also pointed out that the overall mass loss
from the supergiant should be about a factor of 3\,--\,10 smaller than
the values inferred from ultraviolet line studies that assume a
homogeneous wind. This, as well as all other works on the
phenomenology of clumped winds in HMXBs, account only for the presence
of dense wind clumps and ignore the non-monotonic wind velocity.

In this paper we employ  time-dependent hydrodynamic simulations of
non-stationary stellar winds and use the predicted non-monotonic
density and velocity fields to compute the accretion rate. We
deliberately concentrate on the general consequences of stellar wind
clumping for the HMXBs. Multi-D hydrodynamic
models and the processes in the immediate vicinity of the compact
object are beyond the scope of this paper. We consider here systems of
intermediate and small X-ray luminosity, such that the donor
stellar wind at large is not perturbed by the presence of
the compact object and its X-ray emission.

The paper is organized as follows. In Sect.\,\ref{sec:hydro} we
describe the hydrodynamic models used to obtain the time-dependent
wind density and velocity.  The Bondi-Hoyle-Lyttleton accretion
formalism is applied in Sect.\,\ref{sec:hb} to simulate the
resulting time-dependent X-ray luminosity. The photoionization in
the clumped wind is considered in Sect.\,\ref{sec:ksi}.  In
Sect.\,\ref{sec:abs} we apply our 2D model of a clumped stellar wind
to predict the changes in absorption during the orbital motion of the NS.
The discussion is presented in Sect.\,\ref{sec:disc} and conclusions
are drawn in Sect.\,\ref{sec:con}.

\section{Hydrodynamical simulations}
\label{sec:hydro}

We employ the results of hydrodynamical simulations of the line driven
stellar wind presented in \citet{feld1997a,feld1997b}. These
hydrodynamic simulations derive the dynamic and thermal structure of
stellar winds from the basic underlying physical principle, namely the
acceleration of stellar wind by the line scattering of stellar UV
photons (CAK).

The simulations were performed using the smooth source function method
\citep{ow1996}. The line-driven instability (LDI) is handled by a
careful integration of the line-optical depth with a resolution of three
observer-frame frequency points per line Doppler width and by accounting
for non-local couplings in the non-monotonic velocity law of the wind.
The angle integration in the radiative flux is limited to one single ray
which hits the star at ${\approx}0.7 R_\ast$. This accounts with
sufficient accuracy for the finite-disk correction factor.  The pure
hydrodynamics part of the code is a standard van~Leer solver.   The seed
perturbations for unstable growth, in the shape of turbulent variation of
the velocity at a level of roughly one third of the sound speed, are
introduced at the base of the wind. These perturbations have a coherence
time in the friction term of 1.4\,hr, which is not too far off the
acoustic cutoff period of the star at which acoustic perturbations of
the photosphere should accumulate. Further details can be found in
\citet{feld1995,feld1997a,feld1997b}.

The hydrodynamic simulations were performed using stellar parameters of
a typical late O-type supergiant star $\zeta$\,Orionis (O9.7Ib) (see
Table\,\ref{tab:starmod}). A snapshot of a wind structure at some given
moment of time is shown in Fig.\,\ref{fig:rhov}. The variety of
dynamical structures in the instability-induced, line-driven wind
turbulence seems to be similarly intricate to that found in other
turbulent flows in astrophysical MHD settings. There are indications of
the presence of a quasi-continuous hierarchy of density and velocity
structures in the wind.  Dense shells are formed close to the star,
in a first stage of unstable growth. These dense shells have rather
small radial extent, and contain a large fraction of the wind mass. In
the framework of these hydrodynamic models, the colloquial term ``wind
clumps'' refers to these dense shells. The space between the shells is
filled by tenuous and hot gas \citep{feld1997a,feld1997b}.

Further out in the wind, in a second stage of unstable growth,
a somewhat different type of dense structures are formed: the small
``clouds'' that are ``ablated'' from the outer rim of the remaining
mass reservoir at CAK densities, and are accelerated by the stellar
radiation field through the emptied regions, eventually colliding with
the next-outer shell. The X-ray emission from single early type star
winds is explained by these cloud-shell collisions.

\begin{figure}
\centering
\includegraphics[width=\columnwidth]{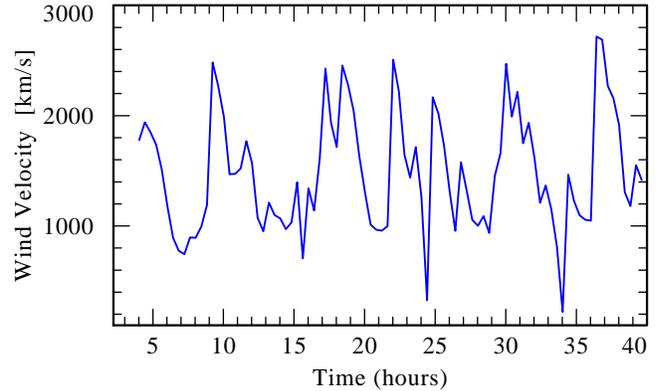}
\caption{The model (see text) wind velocity as function of time at the
radius $a=2.5\,R_\ast$. The amplitude of velocity variations may
exceed order of magnitude in only few hours. }
\label{fig:vt}
\end{figure}

{\changed 
While the \emph{average} velocity field follows the so-called $\beta$-law:
\begin{equation}
 v(r)=v_\infty(1-1/r)^\beta,
\label{eq:vb}
\end{equation}
where $v_\infty$ is the wind terminal speed,  there are strong velocity 
jumps with negative gradients across the shells. }In Figure\,\ref{fig:vt}
we show the wind velocity at some specific radius in the wind as function
of time. The figure illustrates, that relatively close to the
star (we selected a radius $2.5R_\ast$ from the stellar surface) the wind
velocity changes by up to an order of magnitude on a time scale of hours.

\section{Accretion of a non-stationary stellar wind}
\label{sec:hb}
\subsection{Basics of Bondi-Hoyle accretion} 

The following is based on the treatment of \citet{do1973}. They considered a 
neutron star of mass $m_{\rm X}$, traveling at relative speed $v_{\rm
  rel}$ through a gas density $\rho$. If $v_{\rm rel}$ is supersonic,
the gas flows in a manner described by \citet{bh1944}. The mass
accretion rate is given by
\begin{equation}
S_{\rm accr}=\pi \zeta r_{\rm accr}^2 v_{\rm rel} \rho(a,t)
\label{eq:acrrate}
\end{equation}
where  
\begin{equation}
r_{\rm accr}=\frac{2Gm_{\rm X}}{v_{\rm rel}^2}
\label{eq:raccr}
\end{equation}
and the factor $\zeta$ is intended to correct for radiation pressure
and the finite cooling time of gas. In moderately luminous X-ray
sources $\zeta\lsim 1$ \citep{do1973}. 
{\changed For our calculations we adopt $\zeta = 1$.}

In non-stationary stellar winds, the wind density
$\rho(a,t)$ and the wind velocity $v_{\rm w}(a,t)$ depend on time (see
Section\,\ref{sec:hydro}).
Combining equations (\ref{eq:acrrate}) and (\ref{eq:raccr}) the 
accretion rate is
\begin{equation}
S_{\rm accr}=4\pi \zeta \frac{(Gm_{\rm X})^2}{v^3_{\rm rel}}\rho(a,t).
\label{eq:sac}  
\end{equation}
The relative velocity 
\begin{equation}
v^2_{\rm rel}=v^2_{\rm X}(a) + v^2_{\rm w}(a,t), 
\label{eq:vrel}
\end{equation}
where   $v_{\rm w}$ is the stellar wind velocity at the neutron star orbital 
separation $a$, and the orbital velocity of the neutron star is given by 
\begin{equation}
v^2_{\rm X} \approx \frac{GM_\ast}{a}. 
\label{eq:vx}
\end{equation}
The X-ray luminosity of an accreting neutron star is then 
\begin{equation}
  L^{\rm d}_{\rm X}(a,t)=\eta S_{\rm accr} c^2,
\label{eq:lx}
\end{equation}
where $c$ is the speed of light, and $\eta$ is a
constant depending on the exact physics of accretion and typically taken
to be $\eta\sim 0.1$. 

%

\subsection{Synthetic accretion light curves}

The time dependent hydrodynamic simulations described in 
section\,\ref{sec:hydro} provide absolute values for the wind density
and velocity as functions of radius and time 
for a star with parameters as shown in Table\,\ref{tab:starmod}. We use these 
density and velocity to compute synthetic X-ray light curves according
to Eqs.\,(\ref{eq:sac})--(\ref{eq:lx}).

\begin{figure*}
\centering
\includegraphics[width=12cm]{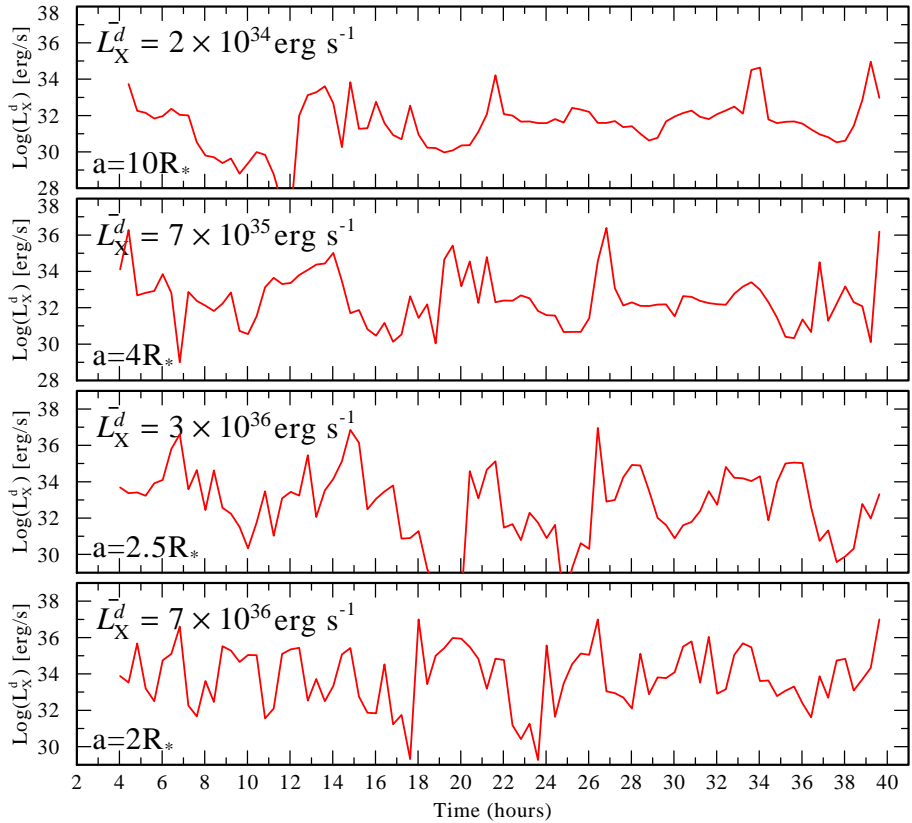}
\caption{Synthetic X-ray light curves for Bondi-Hoyle accretion of 
a non-stationary stellar wind onto a neutron star with 
$m_{\rm X}=1.4\,M_\odot$. 
The supergiant donor parameters are given in Table\,\ref{tab:starmod}. 
From top to bottom panel, the orbital separation $a$ is decreasing, 
as indicated in the lower left corner of each panel. The average 
X-ray luminosity is indicated in the upper left corner. The relative 
variability $\sigma_L/\bar{L}_{\rm X}^{\rm d}=4.8,\,4.9,\,4.9,\,4.6$ 
from top to bottom panel. }
\label{fig:slc}
\end{figure*}

Adopting a mass of the compact object $m_{\rm X}=1.4\,M_\odot$ and an
efficiency parameter $\eta=0.1$, we generated simulated light curves 
for different values of the orbital separation $a$. The resulting
predicted variations in X-ray luminosity $L_{\rm X}^{\rm d}$ are shown in
Fig.\,\ref{fig:slc}.

The synthetic light curves are highly variable, with X-ray luminosity
changing by up to eight orders of magnitude! This strong variability is 
a consequence of the highly variable density and velocity in the stellar 
wind. Specifically we would like to emphasize that the shocks and
corresponding jumps in the velocity with $\delta v \sim 
1000$\,km\,s$^{-1}$ are responsible for the predicted strong variability, 
since the accretion rate scales with $\propto v^{-3}$. Thus models which
account only for the density fluctuations in the winds
\citep[e.g.][]{wal2007,duc2009} may underestimate the
resulting variability. 

It is interesting to note, that the character of the light curves cannot
be characterized as steady quiescent emission and some strong flares on
top it. Reflecting the nature of non-stationary inhomogeneous stellar
winds the accretion light curves are truly stochastic.

\begin{figure}
\centering
\includegraphics[width=0.99\columnwidth]{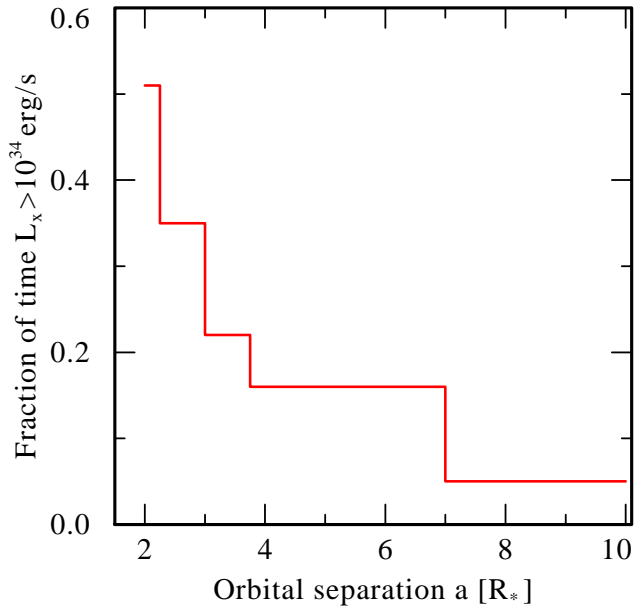}
\caption{The fraction of time the synthetic accretion light curve is
 above  the level $10^{34}$\,erg\,s$^{-1}$ in dependence on the orbital 
separation $a$.}
\label{fig:frac}
\end{figure}

The X-ray observations of real celestial HMXBs are sensitive to the
X-ray flux only about some threshold. In an arbitrary way we choose this
threshold corresponding to the X-ray luminosity
$10^{34}$\,erg\,s$^{-1}$.  Figure\,\ref{fig:frac} shows the fraction of
time the luminosity of model source is above this threshold.  
It demonstrates that the fraction of time the source
is X-ray bright {\nchanged decreases sharply with increasing } orbital separation.
In HMXBs with larger orbital separation, the periods of highest
accretion are significantly less frequent.

%

We calculate the average X-ray luminosity resulting from direct accretion,
$\bar{L}_{\rm X}^{\rm d}$, the standard deviation $\sigma_L$, and the
relative variability $\sigma_L/\bar{L}_{\rm X}^{\rm d}$. Our models
show that there is no apparent dependence of the relative variability
on orbital separation within the parameter space we have chosen. This
reflects the stochastic nature of stellar wind even relatively far
from the stellar core.

\section{The photoionization parameter in the clumped wind of HMXBs}
\label{sec:ksi}

The total X-ray emission in HMXB results from direct accretion which
we considered in the previous section as well as from the photoionized
fraction of stellar wind. Due to the highly variable accretion rate, the
ionizing source is  highly variable (see Fig.\,\ref{fig:slc}). Moreover,
the ambient stellar wind is highly inhomogeneous (see
Fig.\,\ref{fig:rhov}). In this section we investigate how the ionization
parameter is changed in a clumped wind compared to a smooth medium --
the latter is commonly adopted in the computations of spectra from
photoionized  plasma.

To study these effects we adopt a statistical approach, assuming a large
number of clumps. The use of a statistical approach is convenient for at
least two reasons. First, the non-stationary hydrodynamic simulations we
presented in the previous section are 1D, and while they are well suited
to compute the luminosity of the accreting point source, we need
additional information to study spatially extended photoionized regions.
Second, there are sophisticated photoionization codes, such as
\emph{xstar} \citep{kal2001} that account realistically for atomic
physics of photoionized gases. Using a statistical treatment of the
clumped stellar wind, we attempt to find correction factors that would
allow to adjust the input parameters for a photoionization code in order
to reflect the wind inhomogeneity.

{\changed  The fundamentals of interaction of X-ray sources with ambient
stellar winds are developed in \citet{tts1969}, \citet{ts1969}.
\citet{hmc1977}  considered a model of a constant X-ray source immersed
in a stellar wind from its companion star. The stellar wind model was as
predicted by the then recent CAK theory. These works consider a nebula
gas of constant density $N$  (or smooth density gradient) illuminated by
a  constant source of X-ray radiation. The ratio $N_{\rm e}/N$ is
determined in  each case from charge neutrality. Adopting
\citet{tts1969}  notations, neglecting absorption one may take for the
mean primary photon  intensity at frequency $\nu$ at a distance $r$ from
the source
\begin{equation}
J(\nu){\rm d}\nu=\frac{L}{4\pi r^2}f(\nu,kT_{\rm X}), 
\label{eq:ju}
\end{equation}
where function $f(\nu,kT_{\rm X})$ describes the spectral shape of the 
X-ray radiation. The equation of ionization equilibrium for a homogeneous 
medium is 
\begin{equation}
N(Z,z+1)N_{\rm e}\alpha(T_{\rm e})=
N(Z,z)\int J(\nu)\sigma(\nu){\rm d}\nu+N_{\rm e}C(T_{\rm e}),
\label{eq:smie}
\end{equation}
where $C$ is the collisional ionization rate and $\sigma$ is the
photo-ionization cross-section for the changing ionic charge $z
\rightarrow (z+1)$ of ion with  nuclear charge $Z$; $\alpha$ is the
total coefficient for recombination for $(z+1) \rightarrow z$. 

As follows from Eq.\,(\ref{eq:ju}) and Eq.\,(\ref{eq:smie})  for an
optically thin gas in local ionization and thermal balance,  irradiated
by a constant point source of X-rays of given spectral shape,  only a
single photoionization parameter must be specified in order to
determine  the ionization state and temperature of any local region of
the gas: $ \xi \equiv {L_{\rm X}}/{N a^2}$, 
where $N$ is the local atomic number density of the gas at the orbit
of the X-ray source, and $a$ is the distance from the center of the
primary to the X-ray source. 

In the medium where the density is inhomogeneous, Eq.\,(\ref{eq:smie}) 
shall be rewritten as 

\begin{equation}
\langle N(Z,z+1)N_{\rm e}\rangle \alpha(T_{\rm e})=
\langle N(Z,z)\rangle\int J(\nu)\sigma(\nu){\rm d}\nu+
\langle N_{\rm e}\rangle C(T_{\rm e}),
\label{eq:inhie}
\end{equation}
where the averaging over some local volume is represented by the
angular  brackets. Considering Eq.\,(\ref{eq:inhie}) it is clear,
that, in general,  the ionization parameter $\xi$ is not the same as in
smooth density case.   It is also clear, that in the medium with density
fluctuations, the number of  recombinations which is proportional to the
density squared is enhanced compared  to the number of ionizations which is
proportional to the density. Therefore, in the  clumped stellar wind,
the photoionization parameter must be reduced compared  to the smooth
wind. Moreover, the size of the photoionized region will be reduced in a 
clumped wind compared to the smooth wind of the same mass-loss rate.

To describe the density fluctuations we follow \citet{allen1973} and 
define the wind inhomogeneity parameter
\begin{equation}
{\cal X} \equiv \frac{\langle N^2 \rangle}{\langle N\rangle^2}.
\label{eq:x}
\end{equation}
Then the root mean square (rms) local density is $N_{\rm rms}= \langle N
\rangle \sqrt{\cal X}$. The rms density describes in a statistical 
sense the magnitude of density variations.

Assuming that there are no dramatic changes in the ionization state
over  some local volume, we can write
\begin{equation}
 \langle N(Z,z+1)N_{\rm e}\rangle\approx 
{\cal X}\langle N(Z,z+1)\rangle \langle N_{\rm e}\rangle,
\label{eq:xin}
\end{equation}
In this case, the photoionization parameter for a clumped stellar wind 
is
\begin{equation}
\xi \equiv \frac{L_{\rm X}}{{\cal X} \langle N \rangle a^2},
\label{eq:inksi}
\end{equation}
i.e.\ it is reduced by factor ${\cal X}$ compared to the smooth wind case. 

The particle density  averaged over some unite volume at distance 
$r$, $\langle N \rangle$ is defined by the continuity equation:
\begin{equation}
\langle N \rangle = \frac{\dot{M}}{4\pi\mu m_{\rm H} v(r)r^2},
\label{eq:avn} 
\end{equation}
\noindent 
where $\mu$ is the mean molecular weight, $m_{\rm H}$ is the atomic mass
unit, and $v(r)$ is a stationary velocity law.

The simplest structure of a clumped wind is a two phase medium where
all  matter is clumped with dense clumps occupying only a small volume
compared to the interclump medium filling the large volume. The
interclump medium can be assumed void or filled with  tenuous gas. In
this  case a volume filling factor  $f_{\rm V}$ can be defined as the
fraction of total volume $V_{\rm tot}$ that is filled by clumps: $f_{\rm
V} \equiv V_{\rm cl}/V_{\rm tot}$.  If the interclump medium is void
${\cal X}=f_{\rm V}^{-1}$.  The clump  density in such two phase medium
is $N_{\rm c}={\cal X}\langle N\rangle$.   If the interclump medium is
not void, its particle number  density can be approximated as: 
\begin{equation} 
N_{\rm ic}\approx (1-{\cal X}f_{\rm V}) \langle N \rangle.  
\label{eq:pard}  
\end{equation} 

\subsection{On the clump optical depth}
\label{sec:thicl}

Equation\,(\ref{eq:ju}) neglects the absorption of ionizing radiation.
{\nchanged  This condition may not be valid when clumps are optically thick
for the ionizing radiation. In this section we investigate the restrictions 
on the clump parameters imposed by this condition.} To do so we estimate 
the geometrical size a clump which has 
optical depth unity. Assume for simplicity that the interclump medium is void, 
so that $f_{\rm V}={\cal X}^{-1}$, the optical depth of an isotropic 
clump (i.e., with the same dimensions $d_{\rm c}$ in 3D)  
for the incident radiation at wavelength $\lambda$ is
\begin{equation}
\tau_{\rm c}(\lambda)= 
\rho_{\rm w}\kappa_{\lambda} f_{\rm V}^{-1} d_{\rm c}\Rstar,
\label{eq:tc}
\end{equation}
where $R_\ast$\,[cm] is the stellar radius, $d_{\rm c}$ is in the units
of $R_\ast$ and 
$\kappa_\lambda$ [cm$^2$\,g$^{-1}$] is  the mass absorption coefficient.
The density of the cool wind ($\rho_{\rm w}$) can be defined from the 
continuity equation.

The clump dimension $d_{\rm c}$ can be derived from 
Eq.\,(\ref{eq:tc}). Adopting  $\tau_{\rm c}(\lambda)=1$: 
\begin{equation}
d_{\rm c}^{\tau=1}=\frac{f_{\rm V}}{\tau_\ast(\lambda)}
\cdot\left(1-\frac{1}{r}\right)^{\beta}r^2. 
\label{eq:dc1}
\end{equation}
where $\tau_\ast$ is a parameter and a $\beta$-law for the velocity is
assumed.  If at some radius $r$, the clump size is larger than 
$d_{\rm clump}^{\tau=1}$, such a clump is no longer optically thin for X-ray
radiation at $\lambda$. 
The parameter $\tau_\ast(\lambda)$ can be conveniently expressed as
\begin{equation}
\tau_\ast(\lambda)\,\approx\,
722\frac{\mdot_{-6}}{v_\infty {\cal R}_\ast}\cdot\kappa_\lambda,
\label{eq:722}
\end{equation}
where $\mdot_{-6}$ is the mass-loss rate in units $10^{-6}$\,\myr,
$v_\infty$ is the terminal velocity in [km\,s$^{-1}$], and ${\cal
R_\ast}=\Rstar/R_\odot$ \citep{osk2011}.  

The K-shell opacity of stellar wind is a strong function of  wavelength,
and  can be taken as  $\kappa_\lambda\approx \kappa_0 \lambda^{\gamma}$,
with $\gamma$ in the range 2--3, and $\kappa_0$ a constant that depends
on abundances \citep{baum1992,ignace1999}. It follows from
Eqs.\,(\ref{eq:dc1}) and (\ref{eq:722}) that the same clump can be
optically thin at shorter wavelengths and  optically thick at longer
wavelengths.

Realistic values of the mass-absorption coefficient have to be
computed using non-LTE model stellar atmosphere codes, such as, e.g.,
PoWR \citep{hg2004}. The values of $\kappa_{\lambda}$ for a sample of
O-stars are given in e.g.\,\citet{osk2006}. 

The SG HMXB are hard X-ray sources. The maximum of spectral energy 
distribution in Vela\,X-1 is at $\approx 4$\,\AA\
\citep[e.g.][]{dor2011}. Using parameters of $\zeta$\,Ori and assuming 
$f_{\rm V}=0.1$, at $a=2\,R_\ast$,  $d_{\rm clump}^{\tau=1}(\zeta\,{\rm
Ori})\approx 20\kappa_{\lambda}^{-1}$. I.e.\, at the distance
$2\,R_\ast$ from the stellar core, only geometrically  large clumps with
diameter larger than $\sim 0.5\,R_{\ast}$ are optically thick  for the
ionizing radiation shortwards of 10\,\AA.  

Such a clump size appears rather large, therefore it is probably
reasonable to assume that the wind clumps are optically thin for the
ionizing radiation resulting  from the accretion onto a NS 
\footnote{Note, that the intrinsic X-ray radiation in single 
SG OB stars is thermal and soft, with the maximum of the differential 
emission measure  distribution at $\approx 60$\,\AA\ or even softer
\citep[e.g.][]{raas2008}. Even  geometrically very small clumps with
diameter $10^{-3...-5}\,R_\ast$ are optically thick at these wavelengths. 
Therefore,  generally, in the inner wind regions of {\em single} SG OB 
stars the optically thin clump approximation for the intrinsic X-ray 
radiation is false.}

\subsection{Optically thin clumps}
\label{sec:thin}

As shown above, the wind clumps are most probably thin 
for the hard X-ray emission resulting from the accretion on a compact object.  
In this case, the ``microclumping'' approximation can be used. If the
mean primary photon intensity is described by Eq.\,(\ref{eq:ju}), then 
in a two phase medium where ${\cal X}=f_{\rm V}^{-1}$, the photoionization 
parameter can be written as 
\begin{equation}
\xi = f_{\rm V} \cdot \frac {L_{\rm X}}{\langle N \rangle a^2}.
\label{eq:ksiin}
\end{equation}
The meaning of Eq.\,(\ref{eq:ksiin}) is that the ionization parameter 
in a clumped wind is reduced by a factor $f_{\rm V}$ when the average
density at distance $a$ is given by Eq.\,(\ref{eq:avn}). Following 
\citet{hmc1977} the nondimensional variable 
$q\equiv \xi \langle N \rangle a^2 (L_{\rm X}f_{\rm V})^{-1}$ 
describes the shape of the surfaces for which $\xi\,=$\,constant.
In clumped stellar wind, the parameter $q$ will be larger than in 
a smooth wind. Considering the case of constant wind velocity, it 
would mean that the radius of spherical photoionized region is 
smaller in clumped wind compared to the smooth wind. This is easy 
to understand recalling that the number of recombination is enhanced in 
clumped wind compared to smooth wind, while the number of ionizations
remains the same.

Over the  last decade there have been numerous studies to establish the
degree of wind clumping in OB supergiants. \citet{puls2006} analyzed a
sample of 19 Galactic O-type supergiants/giants, covering spectral types
O3 to O9.5 using multiwavelength observations in radio, IR, and optical.
They used the conventional assumption of the interclump medium being
void. It was  found that {\changed the upper limit on the} volume
filling factor in the  innermost region ( $r \la 2~ R_\star$) of stars
with H$_\alpha$ in  emission is about $0.25$. It is likely that the
winds of donor stars in  HMXBs have similar properties. In this case,
the ionization parameter  should be reduced by a minimum factor of four 
to account for wind clumping.

\citet{ful2006} analyzed 40 Galactic O-type stars by fitting stellar
wind profiles to observations of the P\,{\sc v} resonance doublet in the
UV. The found that to reconcile mass-loss rates empirically inferred
from the analysis of UV and radio/optical spectra, the volume filling 
factors must be $f_{\rm V} \lsim 0.01$. Similar conclusions were reached
by \citet{bour2005} who found $f_{\rm V} \approx 0.04$ in an O-type
supergiant and $f_{\rm V} \approx 0.02$ in an O-type dwarf. If these
clumping factors were realistic, the ionization parameter $\xi$ would
have to be reduced by factors of 25\,...\,100. \citet{osk2007} showed
that correctly accounting for the wind porosity in the analysis of UV
resonant lines eliminates the need for very small volume filling factors
and that realistic values are rather $f_{\rm V} \sim 0.1$. 
{\changed \citet{sun2011} also used the hydrodynamic models 
of \citet{feld1997a,feld1997b} and found that in these models the 
typical clumping factors are $f_{\rm V}^{-1}\sim 10$.  In this case,
in the studies of the X-ray spectra from photoionized gas, the ionization 
parameter $\xi$ should be reduced by a factor $\lsim 10$ in clumped wind 
models compared to the unclumped ones.}

\subsection{Clumps of arbitrary optical depth}
\label{sec:thick}

While it is likely that the wind clumps are optically thin for the hard
ionizing  radiation,  (see Section\,\ref{sec:thicl}), for completeness
in  this section we consider a case of optically thick clumps.  
Accounting for the  clumps of arbitrary optical depth in
stellar atmosphere  models is sometimes referred to as
``macroclumping''. If the optically thick structures, $\tau_{\rm c}\msim
1$, are  present in the wind, the mean primary photon intensity of
ionizing radiation $J_{\nu}$  must be corrected for the absorption in
the clumps
\begin{equation}
J_{\nu}=\frac{L_{\rm X}}{4\pi r^2}\cdot f(kT)\cdot 
{\rm e}^{-\bar{\tau}(r)},
\label{eq:jnu}
\end{equation}
where $f(kT)$ is a function of atomic parameters and temperature that
describes the spectrum of the incident ionizing radiation,  and
$\bar{\tau}$ is the effective optical depth of the clumped medium
\citep{feld2003}.  

At a distance $R_{\rm X}$ from  the source of ionizing radiation the
effective optical depth is 
\begin{equation}
\bar\tau=\int_0^{R_{\rm X}} n_{\rm C} \sigma_{\rm C} (1-{\rm e}^{-\tau_{\rm c}}) 
{\rm d}r_{\rm x},
\label{eq:tint}
\end{equation}
where $n_{\rm C}(r)\propto v^{-1}(r) r^{-2}$ is the average number of
clumps within a unit volume, {\nchanged and $\sigma_{\rm C}$ is the clump
geometrical cross-section \citep{osk2008,osk2011}}. Note that the stellar wind
parameters $\dot M$ and $\kappa_\lambda$ appear in the expression for
$\bar\tau$ only indirectly via $\tau_{\rm c}$ given by
Eq.\,(\ref{eq:tc}).  Besides stellar wind parameters, the effective optical 
depth also depends on the geometrical  distribution of the clumps 
via $n_{\rm C}$ and $\sigma_{\rm C}$. The clump cross-section is different for 
the different clump geometries, e.g.\ for the spherical clumps 
$\sigma_{\rm C}\propto d^2_{\rm C}$, while for the oblate clumps the 
cross-section depends on the clump orientation. In the limit of $\tau_{\rm C}\gg 1$, 
the effective optical depth is fully determined by the geometrical 
distribution  of clumps \citep{feld2003, osk2004}. 

Now we would like to address the question of ionization balance in the
wind where the largest portion of the matter is in form of optically
thick clumps. {\changed The X-ray emission from such medium will have two 
components: X-rays due to scattering of radiation within the clumps, and the 
X-rays from the photoionized interclump medium.} The scattering of photons 
in such a medium was considered by \citet{sul2003}. Here, we consider the 
ionization of the interclump medium {\nchanged and neglect the diffuse radiation}. 
Using Eq.\,(\ref{eq:pard}) for the particle number density in the interclump 
medium, the ionization parameter for the tenuous  interclump gas becomes
\begin{equation}
\xi_{\rm ic}=\frac{L_{\rm X}}{\langle N \rangle a^2}\cdot
\frac{{\rm e}^{-\bar{\tau}}}{1-{\cal X}f_{\rm V}}.
\label{eq:ksithick}
\end{equation}
When the bulk of stellar wind mass is in the form of a few dense optically
thick  clumps, the interclump medium can be very tenuous 
${\cal X} \rightarrow f_{\rm V}^{-1}$. {\changed In this case 
the effective optical depth can be very small (i.e.\ the wind is very 
porous). As follows from Eq.\,(\ref{eq:ksithick}), in this case the 
ionization parameter in the tenuous interclump gas can be very high,}
much higher than would be expected for the smooth  wind of the same
mass-loss rate. 

We conclude that clumping strongly affects the ionization balance.  A
higher degree of ionization in the less dense interclump gas within the
photoionized region can be expected. The size of the photoionized region
will be much larger in case of optically thick clumps compared to 
a smooth wind.
}

\section{Absorption of X-ray radiation in a clumped stellar wind}
\label{sec:abs}

\begin{figure}
\centering
\includegraphics[width=\columnwidth]{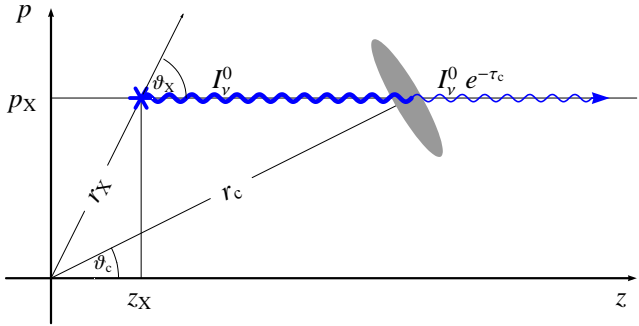}
\caption{Sketch of the coordinate system for the fragmented wind model.  
X-ray radiation with intensity $I^0_\nu$ is emitted at the location with 
coordinates $(p_{\rm i}, z_{\rm i})$. The cool wind consists of clumps  
that are randomly distributed. For clarity, only one clump (shaded area) 
is shown in the sketch. The X-ray radiation propagating towards the observer 
at $z \rightarrow \infty $ has to cross a clump. The optical depth across 
the clump for the ray in $z$ direction is $\tau_{\rm c}$, reducing the 
line intensity $I^0_\nu $ by a factor $\exp (-\tau_{\rm c})$.}
\label{fig:corcl}
\end{figure}

The X-ray photons emitted either in close vicinity to the NS, or farther
out in the photoionized portion of the stellar wind, suffer the K-shell
continuum absorption in the cool component of the stellar wind as they
propagate towards the observer.  {\changed As discussed in 
Sect.\,\ref{sec:thicl} the wind clumps are, most likely, optically thin 
for the radiation at the short wavelengths (shortward of $\approx 8$\,\AA). 
On the other  hand, the wind clumps or larger scale wind structures 
can be optically thick for the softer X-rays. These optically thick 
inhomogeneities can be traced observationally as changes in the absorbing 
column density on the time scale comparable to the orbital period. 
Such variability of absorbing column is indeed detected \citep[e.g.][]{furst2011}. 
If the stellar wind of donor star were smooth, the absorbing column would 
smoothly increase from the moment when the NS is at the periastron to 
when it is in the apastron.

In this section we investigate the K-shell continuous X-ray absorption 
in stellar wind. Our goal is to model the absorption of monochromatic 
X-ray flux emerging from the NS in dependence on orbital phase. We employ 
a  2D model of a stochastic stellar wind by \citet{osk2004}, where the model 
details are fully described. Here we only briefly introduce its basic features.}
The model assumes the structure and physical conditions in the wind
according to the hydrodynamic simulations by \citet{feld1997a,
feld1997b}.  {\changed The emission is decoupled from absorption, 
which allows for the formal solution of the radiative transfer equation.
In order to take advantage of the rotational symmetry of our model around 
the z axis, cylindrical coordinates $(p,z,\varphi)$ are used.
Along each line of sight towards an X-ray emitting region located at 
spherical coordinates $(r_i,\vartheta_i)$,  
the emergent monochromatic X-ray intensity $I^+(p,z,\varphi)$ is reduced by 
a factor $\exp{(-\tau(p,z,\varphi))}$: 
$I^+(p,z,\varphi)=I^0(p,z,\varphi)\exp{(-\tau(p,z,\varphi))}$. 
The optical depth $\tau(p,z,\varphi)$ is given by the summation of optical 
depths of all clumps on the line of sign. The geometry of 
the model is shown in Fig.\,\ref{fig:corcl}.  }
Due to the lack of synchronization between
different radial directions, the fragments probably have rather small
lateral extent. Despite the expansion due to the internal pressure, the
cool fragments are maintained to distances of hundreds of the stellar
radius, before they gradually dissolve into a homogeneous outflow.  Our
2-D stochastic wind model is designed to describe in a generic way the
structures and physical conditions of the inhomogeneous stellar wind:

\begin{figure}
\centering
\includegraphics[width=\columnwidth]{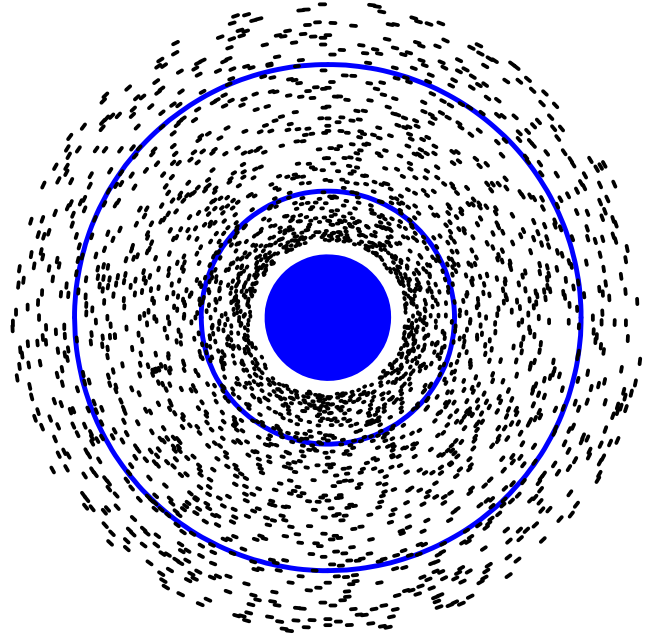}
\caption{One random trial of our 2-D model where the clumps are
  randomly distributed in radius (uniformly between 1.2 and 316
  $R_\ast $). On average there are 400 clumps along each radial
  direction (see text for details).  Only the inner part of the wind
  between 1$R_\ast$ and 5$R_\ast$ is shown. The smooth wind is
  in-between 1$R_\ast$ and 1.2$R_\ast$. The filled circle represents the
  stellar core; the outer circles are at 2$R_\ast$ and 4$R_\ast$ to
  scale. This random model was used to calculate optical depths shown
  in Fig.,\ref{fig:tau19} and Fig.,\ref{fig:tau8}.}
\label{fig:windsketch}
\end{figure}

\smallskip\noindent
1. the flow is spherically symmetric in a statistical sense and
propagates according to a $\beta$ velocity law. The mass-flux and total 
radial optical depth in each radial direction is preserved.

\smallskip\noindent
2.  all clumps cover the same solid angle as seen from the center of
the star, and preserve this angle during their propagation.

\smallskip\noindent
3.  we consider two different shapes of clumps, the spherically symmetric 
clumps (``balls'') and radially compressed clumps (``pancakes'').

\smallskip\noindent    
4.  to simulate stochastic X-ray emission we model X-rays as  discrete
random flashes of X-ray light along the NS orbit. There is no self-absorption 
by emitting material and no re-emission of X-rays which became absorbed 
by cool fragments. 

\smallskip\noindent    
5. {\changed we consider an edge on system, i.e. \ the observer is located in 
the orbital plane of the NS }

\begin{figure}
\centering
\includegraphics[width=\columnwidth]{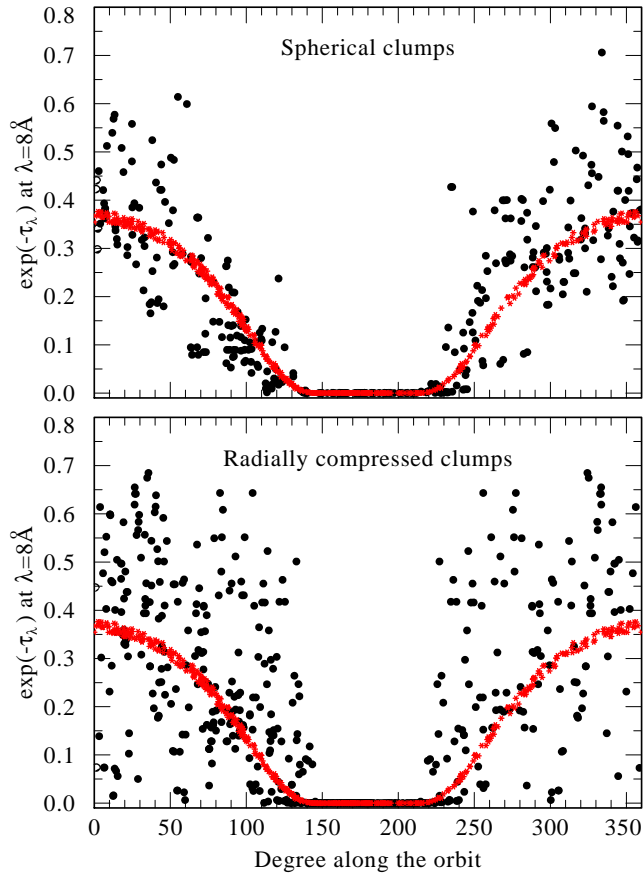}
\caption{The variation of the factor $\exp{(-\tau_{\lambda})}$ in dependence 
on orbital phase for the X-ray radiation at 8\,\AA\ (1.55\,keV) emitted by a 
source located at $\approx 2\,R_\ast$. The horizontal
  axis shows the orbital location of the X-ray source, $\vartheta_{\rm i}$. 
At $0\degr$ the source is in conjunction, while at $180\degr$ it is totally
  eclipsed.  The red curve shows the changes of $\exp{(-\tau_{\lambda})}$ in a smooth
  wind. The parameters of the wind are given in 
  Table\,\protect{\ref{tab:starmod}}. A velocity law
  with exponent $\beta=0.8$ is assumed. The stochastic wind model is
  calculated using the random trial shown in
  Fig.\,\ref{fig:windsketch}.  \emph{Upper panel:} 
  the wind model where the clumps are assumed to be spherically symmetric 
(``\emph{balls}'').  \emph{Lower panel:}  
  the wind model where the clumps are assumed to be flattened in the radial 
direction  (``\emph{pancakes}'').} 
\label{fig:tau8} 
\end{figure}

We neglect the density fluctuations which are likely to result in the NS
wake.  The hydrodynamic simulations of wind structure that  predict the 
existence of a wake were performed for smooth winds
\citep{blond1990}. It is not not obvious how accounting for wind clumping  
can affect the results of hydrodynamic simulations and the shape and
density in the wake. Moreover, the presence of the wake should
introduce some regular pattern in the absorption of X-rays along the
orbital motion of the NS, while our goal is to investigate the basic
effects of stochastic  winds on the absorption of X-rays.  

We also neglect the changes in the ionization structure of the wind due
to the X-ray photoionization close to the NS.   As discussed in
Section\,\ref{sec:ksi}, the clumping is likely to reduce the size of
photoionization zone. The optical depth of the clumps within the 
photoionized portion of the wind may be different from the rest of the
wind, however this effect may be small on average due to the difference
in the size of the general wind, and its photoionized portion.

Figure\,\ref{fig:windsketch} shows a snapshot of the inner wind  structure, 
between 1\,$R_\ast$ and 5\,$R_\ast$ (in the simulations clumping is maintained 
up to 316\,$R_\ast$). We assume the lateral extend of clumps to be $1\degr$.
To study how the wind optical depth changes during the orbital motion of
the NS, we assumed the donor stellar parameters as listed in
Table\,\ref{tab:starmod}. The clump number density obeys the  continuity
equation, therefore the clump density is higher in the innermost part of
the wind and gradually decreases outwards. To find the optical depths
towards the observer at different orbital location, we sum up the
contributions from all the clumps along the line of sign.

The mass-flux in each direction is preserved in our model. 
To calculate the optical depth of each clump (in total
there are 144\,000 clumps in the wind model, part of which is shown in 
Fig.\,\ref{fig:windsketch}) we assume that a clump mass equals to the
mass of the smooth wind confined between to random subsequent radii in
the wind $r_{\rm a}$ and $r_{\rm b}$ (see \citet{osk2004} for details). 
For a star with the parameters shown in Table\,1 and solar abundances,
the mass-absorption coefficient $\kappa_\lambda$ is 
$\approx 40$\,cm$^2$\,g$^{-1}$ at $\lambda=8$\,\AA\ and 
$\approx 170$\,cm$^2$\,g$^{-1}$ at $\lambda=19$\,\AA\ \citep{osk2006}. 

Little is known about the geometrical shape of the clumps. The
comparison with the observed X-ray emission line profiles in spectra
of single O-stars indicate that the clumps {\nchanged can be} flattened
in radial direction \citep{osk2006}.   The shape of clumps plays 
an important role in the transfer of X-ray radiation through the wind: 
in case of flattened clumps the wind opacity becomes anisotropic, while it
remains isotropic (the same as for the smooth wind opacity albeit smaller) 
in case of ball-like clumps  \citep{osk2006}. 

\begin{figure}
\centering
\includegraphics[width=\columnwidth]{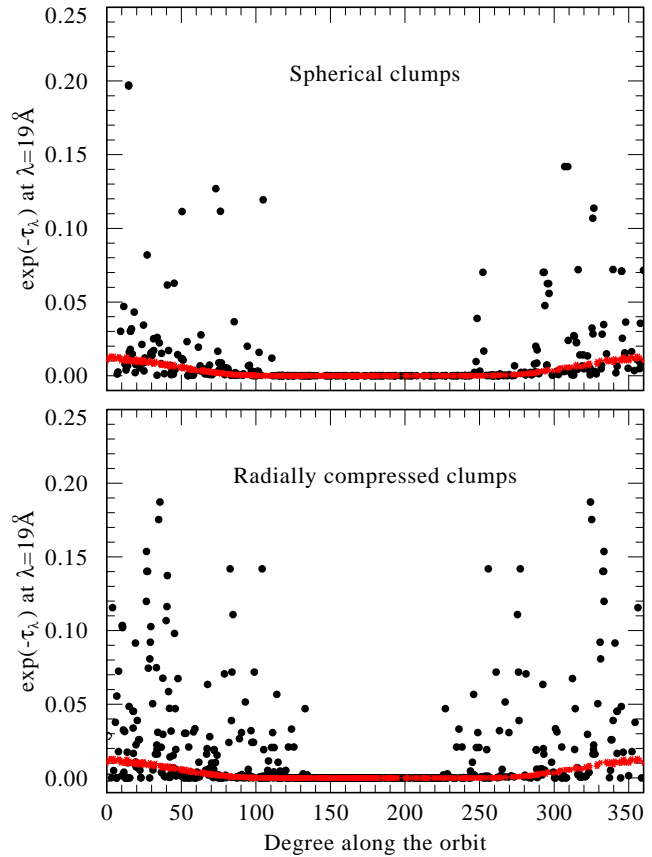}
\caption{The same as in Fig.\,\ref{fig:tau8} but for the radiation 
emitted at 19\,\AA\ (0.65\,keV). }
\label{fig:tau19}
\end{figure}

Figures\,\ref{fig:tau8} and \ref{fig:tau19} show the transmission factor 
$\exp{(-\tau_\lambda)}$ for radiation at 8\,\AA\  and 19\,\AA\ (1.55\,keV and 
0.65\,keV, respectively) calculated  assuming different clump geometries and a 
smooth wind. The wind is more transparent for the radiation at shorter 
wavelengths. In case of flattened clumps the changes in optical depth are 
larger, because the clump optical depth depends on the clumps orientation. 

\begin{figure}
\centering
\includegraphics[width=\columnwidth]{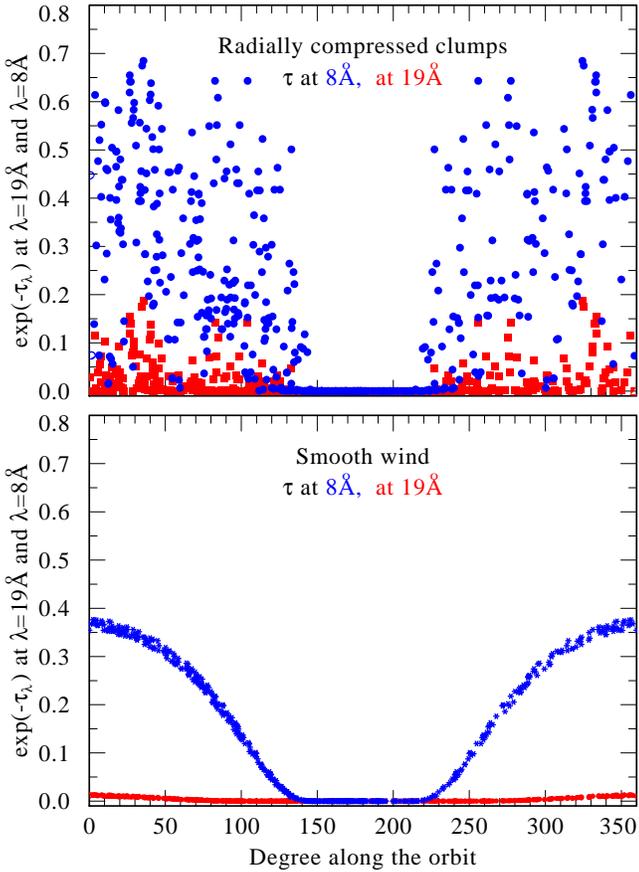}
\caption{The same as in Fig.\,\ref{fig:tau19}. 
\emph{Upper panel:} comparing the $\exp{(-\tau_{\lambda})}$ factor at 
19\,\AA\ (red) and 8\,\AA\ (blue) at different orbital phases for a 
model wind that consists of flattened clumps.  
\emph{Lower panel:} the same as upper panel but for a smooth wind model.}
\label{fig:ecl}
\end{figure}

The changes in the wind absorbing column at different orbital phases 
should result in changes in the X-ray spectra. Indeed, the strong 
variations of absorbing column density at different orbital phases 
are often deduced from the observed spectra \citep[e.g][]{furst2010,boz2011}. 

The K-shell absorption in stellar wind is strongly wavelength
dependent, changing by a two orders of magnitude across the
0.1-12\,keV X-ray band. This is reflected in smaller variations in the
optical depths predicted for the 8\,\AA\ radiation (where the wind
opacity is smaller, Fig.\,\ref{fig:tau8}) compared to the large
optical depth variations at 19\,\AA\ (where the wind opacity is
larger, Fig.\,\ref{fig:tau19}).  The attenuation of X-ray radiation in
a clumped wind has a weaker dependence on the wavelength than in a
smooth wind \citep{osk2004}. This prediction holds also for the
HMXB. In Fig.\,\ref{fig:ecl} we compare the duration of the X-ray
eclipse at two different wavelengths for clumped and smooth wind
models.  In a smooth wind, the eclipse at longer wavelength starts
earlier, and lasts longer, while at shorter wavelength the eclipse is
more narrow. On the other hand, in a clumped wind model, the
difference between eclipse durations is less pronounced. 
{\nchanged Even in the phases close to the eclipse, optical depths 
may occasionally be as low as in the phases far from the eclipse.}

\section{Discussion}
\label{sec:disc}

It emerged from our analysis that the stellar wind clumping has
drastic consequences for the phenomenology of SG HMXBs. The effect of
wind structure on the accretion rate is profound. The hydrodynamic
simulations predict wind structure consisting of strong velocity and
density jumps. Since the Bondi-Hoyle-Lyttleton accretion rate depends
on {\em both} velocity and density fluctuations, it strongly varies
with time. Combining  the Bondi-Hoyle model of accretion with 
hydrodynamic wind simulations we computed synthetic X-ray light curves  
for the different orbital separations (see Fig.\,\ref{fig:slc}).

Averaged over time model X-ray luminosity is in general agreement with
observations.  The moderate reduction of $\dot{M}$ by factor 2\,...\,3
compared to the unclumped wind models explains well the optical and
the UV spectra of single stars, as well as average X-ray luminosity
resulting from wind accretion in HMXBs.

The application of the Bondi-Hoyle-Lyttleton approximation 
for the accretion rate predicts very strong X-ray
variability: the X-ray luminosity changes by many orders of magnitude
on the time scale of hours as a direct consequence of the strong
density and velocity jumps in the accretion flow.  The effect of the
non-stationary wind velocity on the accretion rate was overlooked
previously. Neglecting this effect may lead to erroneous estimates of
clump mass and size from the total flux and duration of the X-ray
flare. We emphasize that the strong velocity jumps in the stellar
winds lead to strong variability of accretion X-ray luminosity.

The detailed comparison of model light-curves and the observations
will be a focus of our forthcoming study. However, it appears safe to
conclude, that the synthetic light-curves do not agree well with the
observations over-predicting the amplitude of variability. Given our
simple approximation for the accretion this is not surprising. The
detailed hydrodynamic studies that wind accretion  is a highly complex 
process. E.g.\, when there is a density and velocity gradient in the flow, 
a transfer of angular momentum can become possible.  The hydrodynamics 
of Bondi-Hoyle-Lyttleton accretion in a medium with continuous density 
and velocity gradients was considered by \citet{ruf1997,ruf1999} and 
\citet{fog2005}. To our knowledge, up to now there were no studies of 
accretions from a medium with strong velocity and density jumps. We 
speculate that the detailed physics of the accretion flows regulates 
the temporal behavior of accretion and is pivotal in damping the strong 
variations due to the structured nature of stellar winds.

{\changed Another process that can reduce the variability is the destruction 
of clumps in the vicinity of the NS.}
As an interesting example of clump destruction, \citet{pit2007}
performed a hydrodynamical simulation of wind-wind collision in massive
non-degenerate binary star. The simulations revealed that the clumps
are rapidly destroyed after passing through the confining shocks of
the wind-wind collision region (assuming adiabatic flow in the wind
collision region). Despite large density and temperature fluctuations
in the postshock gas, the overall effect of the interaction is to
smooth the existing structure in the winds. One may speculate that in
conceptually similar fashion, the wind clumps and the wind velocity
gradients are ``smoothed out'' in the strong shock of the accretion
wake in HMXBs.

\citet{furst2010} found that the accreted clump masses derived from
the INTEGRAL data on Vela\,X-1 are on the order of $5\times
10^{19}\,-\,10^{21}$\,g. \citet{boz2011} estimate a mass and a 
radius of a clump responsible for a flare in a SFXT as $\sim
10^{22}$\,g and $\sim 10^{12}$\,cm correspondingly. These are important 
independent estimates of a clump mass that can be compared with what we 
know about clumps in single stars.
\citet{lep1999} found stochastic line profile variations in the form
of narrow ($\approx$100\,km\,s$^{-1}$) emission sub-peaks on top of
emission lines in Wolf-Rayet star winds. They explained their data by
$10^3 \lsim N_{\rm C}\lsim 10^4$ "blobs'' being present at any time in
the line emission region.  In \citet{osk2006} we reproduced the
Chandra observations of O-star X-ray line profiles by adopting a
dynamic picture from \citet{feld1997a,feld1997b} hydrodynamic
simulations --  {\nchanged in each radial direction} a clump is ejected 
once per dynamic time scale $t_{\rm fl} = R_\ast/v_\infty$, which corresponds 
to a radial separation of one stellar radius when $v_\infty$ is reached.
{\nchanged  Assuming that the wind consists of $N_{\rm C}$ clumps, 
an average clump mass $M_{\rm C}$ is 
\begin{equation}
M_{\rm C} \approx 4.4 \times 10^{25}\,
\frac{\dot{M}_{-6} R_\ast}{v_\infty N_{\rm C}}~[{\rm g}]
\end{equation}
Adopting a mass-loss rate $\dot{M}_{-6} =1$\,[$10^{-6}$\myr], 
$v_\infty = 2000$\,[km\,s$^{-1}$], and $R_\ast=10\,R_\odot$} with 
$N_{\rm C}\sim 10^4\,...\,10^5$ clumps in the wind, the average mass of a
clump is $M_{\rm C}\sim 10^{18}\,...\,10^{19}$\,g. These numbers compare very well
with those independently found by \citet{furst2010} from the analysis
of the INTEGRAL light-curves.

In this study we briefly addressed the absorption of X-rays in clumped
stellar wind. We found that the wind optical depth strongly varies
in dependence on orbital phase. We neglect the photoionization
wake. Partly, this can be justified, because not much is known about
the driving of X-ray photoionized clumped winds. Classically, the
works that predict the decrease of driving force in the X-ray
photoionized part of the outflow consider a smooth wind
\citep{macgreg1982, stev1988, feld1996}. In clumped wind the
ionization balance may be altered, e.g.\, the low ionization stage
that can drive stellar wind may be still present within the clumps.
Our investigation of of changing optical depth in the wind around the
compact object orbit demonstrates the potential of high-quality X-ray
observations of SG HMXBs to provide much needed detailed information
about the parameters of wind clumping. E.g.\, the orientation of the
clumps, their distribution, can be studied from the changing absorbing
column in stellar wind, {\changed especially during the X-ray eclipse.}

The important result of our study, is that the ionization parameter is
affected by the wind clumping. This follows from the basic fact that
in the medium with density fluctuations, the recombination process
($\rho^2$-dependent) have priority over the photoionization process
($\rho$-dependent). {\nchanged The clumping of stellar wind is an 
established fact, and the ionization parameter  must be corrected 
for the clumping in the analyses of spectra of photoionized gas in HMXBs.}  
The ionization structure, the temperature, and the size of photoionized 
region is different in the clumped stellar wind compared to a smooth 
wind with the same mass-loss rate.

\section{Conclusions}
\label{sec:con}

1. The results of time dependent hydrodynamical models of stellar winds
predict a highly variable rate of wind accretion onto a compact
object.  The variability is largely due to the strong variations in the
wind velocity (shocks). The highly non-uniform wind density also
contributes to the variability of accretion rate. 

\medskip\noindent 
2. Combining  the Bondi-Hoyle model of accretion with  hydrodynamic
wind simulations predicts variability on stronger levels than commonly
observed. 

\medskip\noindent 
3. The X-ray light curves synthesized under the assumption of direct
accretion predict strong variability of few orders of magnitude.
The relative variability $\sigma_L/\bar{L}^d_x$ has no strong
dependence on the binary orbital separation.

\medskip\noindent 
{\changed 4. The photoionization parameter depends on the properties 
of wind clumping. In optically thin case, the photoionization 
parameter has to be reduced by the factor ${\cal X}$ compared to 
the smooth wind model, where ${\cal X}$ is the wind inhomogeneity parameter.}

\medskip\noindent 5. The absorption of X-rays in a clumped stellar wind
strongly varies with the orbital phase. The degree of variability 
depends on the shape of the wind clumps, and is stronger for {\changed 
radially compressed} clumps.

\medskip\noindent 6. In a clumped wind, the wavelength dependence of
the duration of the soft X-ray eclipse is smaller compared to the
smooth wind models. 

Overall, we conclude that the clumped non-stationary nature of the wind 
from the donor star strongly affects the X-ray light curve and leads 
to strong variability.

\section*{Acknowledgments}   This research has
made use of NASA's Astrophysics Data System Service and the SIMBAD
database, operated at CDS, Strasbourg, France. The paper benefited from 
the useful remarks and suggestions of the referee, J.\,Sundquist, for which 
we are thankful. Funding for this research has been provided by DLR grant 
50\,OR\,1101 (LMO).  The
research presented here benefited from many useful
discussions with J.\,Wilms, I.\,Kreykenbohm, M.\,Hanke. The
hospitality of Karl Remeis-Sternwarte, Universit\"at Erlangen-N\"urnberg,
is greatly appreciated.  LMO is very grateful for the hospitality and
the simulating research environment of the ESA/ESAC Faculty.

\bibliographystyle{mn2e}
\bibliography{hmxblit}

\label{lastpage}
\end{document}